\documentclass[aps,pra,floatfix,superscriptaddress,twocolumn,showpacs,showkeys,10pt]{revtex4-1}
\usepackage{amssymb, amsmath,color,mciteplus,graphicx,subfigure}
\usepackage{calc,epsfig,epstopdf,color,mciteplus,bm,mathrsfs}
\usepackage{times}
\usepackage{float}
\usepackage{lipsum}

\begin{document}

\title{Fusion of atomic W-like states in cavity QED systems}

\author{Cheng-Yun Ding}\email{cyding@aqnu.edu.cn}
\affiliation{School of Mathematics and Physics, Anqing Normal University, Anqing 246133, China}

\author{Wan-Fang Liu}
\affiliation{School of Mathematics and Physics, Anqing Normal University, Anqing 246133, China}

\author{Li-Hua Zhang}\email{zhanglh@aqnu.edu.cn}
\affiliation{School of Electronic Engineering and Intelligent Manufacturing, Anqing Normal University,
Anqing 246133, China}

\date{\today}

\begin{abstract}
It is well-known that maximally entangled GHZ states can achieve perfect teleportation and superdense coding, whereas maximally entangled W states cannot. However, it has been demonstrated that there exists a special class of non-maximally entangled W states, called as \textit{W-like} states, which can overcome this limitation. Therefore, it is of great significance to prepare such W-like states for efficient quantum communication. Here, we propose two kinds of novel and efficient fusion schemes for atomic W-like states based on the large-detuning interactions between several atoms and a single-mode cavity field, with which large-scale atomic $|\mathcal{W}_{N+M-1}\rangle$ and $|\mathcal{W}_{N+M+T-2}\rangle$ states can be prepared, respectively, from two small-scale atomic $|\mathcal{W}_{N}\rangle$ and  $|\mathcal{W}_{M}\rangle$ states and three small-scale atomic $|\mathcal{W}_{N}\rangle$, $|\mathcal{W}_{M}\rangle$ and $|\mathcal{W}_{T}\rangle$ states, by detecting the states of one or two of the fused atoms. Particularly, although the fusion process of our scheme involves particle loss, the corresponding success probability is high and fixed, which may induce high fusion efficiency. Furthermore, through the investigation of the resource cost and feasibility analysis, our protocol is simple and feasible under the current experimental conditions. All these suggest that it provides an alternative strategy for preparing large-scale atomic W-like states for perfect teleportation and superdense coding.
\end{abstract}


\maketitle

\section{Introduction}
Quantum entanglement, as one of the significant physical resources, is a building block for various quantum information processing (QIP) tasks, such as measurement-based quantum computing \cite{PhysRevA.68.022312,walther2005experimental}, quantum communication \cite{RevModPhys.81.865}, quantum key distribution \cite{RevModPhys.74.145}, and so on. Compared with two-particle entangled states, i.e., Bell state \cite{PhysRevLett.23.880}, multi-particle entangled states exhibit different stories due to their more complex entanglement structures and stronger non-locality. Among them, Greenberger-Horne-Zeilinger (GHZ) state \cite{10.1119/1.16243} and W state \cite{PhysRevA.62.062314} are two key and inequivalent multi-particle entangled states. First, they cannot achieve mutual conversion through local operations and classical communication \cite{PhysRevA.62.062314,PhysRevA.94.052309}; Second, the W state has robustness against qubit-loss \cite{RevModPhys.86.419}, that is, when one or more of its entangled particles is missing, the remaining particles are still entangled, while that for the GHZ state is completely separated. This means that if there are particles loss in the quantum communication, the W state is more favorable. On the other hand, in quantum teleportation and superdense coding schemes \cite{PhysRevLett.70.1895,PhysRevLett.69.2881}, as the two most important quantum communication schemes in the early stage, their performance is just the opposite for transmission efficiency. For example, the GHZ state, also known as the prototype GHZ state, can be used to achieve perfect teleportation and superdense encoding with the success probability and fidelity being both $1$ \cite{joo2003quantum,PhysRevA.63.054301}. However, the prototype W state, also i.e., maximally entangled one, cannot achieve perfect manners because of the need to perform non-local unitary operations or measurements to complete entire communication process \cite{gorbachev2003preparation,PhysRevA.74.062320}. So far, the preparation of Bell state, GHZ state, and W state has been investigated in various quantum systems, such as optics \cite{weissflog2024nonlinear,huang2011experimental,grafe2014chip}, superconducting qubits \cite{PhysRevApplied.11.014017,PhysRevLett.127.043604,PhysRevApplied.18.064036}, and cavity QED system \cite{PhysRevLett.87.230404,PhysRevA.65.042102,PhysRevA.73.014302}. In addition, the larger the entanglement scale (number of entangled particles), the more obvious the quantum superiority which exhibited in quantum communication and quantum computing. So, how to prepare large-scale multi-particle entangled states is urgent and necessary.

In the last decade, it is an effective method to fabricate large-scale multi-particle entangled states using quantum state expansion and fusion technology. For large-scale GHZ states and W states, researchers have proposed many fusion and expansion theoretical schemes in optics \cite{yesilyurt2016deterministic,ozdemir2011optical,PhysRevA.87.032331,li2016generating,zang2017generating,li2018preparing}, atomic systems \cite{PhysRevA.94.062315,zang2016deterministic,zang2015generating,ding2018qubit,shao2024utilizing}, cavity-based nitrogen-vacancy centers systems \cite{han2017effective}, etc. These protocols can be roughly divided into two categories, in the condition that whether the particles involved in the fusion or expansion process have been measured and are lost, that is, those of qubit-loss and qubit-loss-free (QLF). It is an intuitive view that the qubit loss will inevitably lead to a decrease in the efficiency of fusion or expansion process and increase its steps, resulting in an overload on the final resource cost. But counter-intuitively, the success probability of fusion scheme with qubit-loss is higher than that of QLF \cite{PhysRevA.87.032331}, so what is the real reason. Very recently, Huang \textit{et al.} \cite{huang2024quantum} answered this question. They proved that any fusion mechanism is essentially a POVM measurement, and pointed out that fusing standard W states into a large-size W state without particle loss is not the most efficient, so they proposed the fusion scheme for fusing W-like states, a class of special non-maximally entangled W ones, into a large-size W state or W-like states. It is worth noting that the W-like state, that is, the non-maximally entangled state (with unequal weights), may have very meaningful applications. As we all know, in quantum storage and quantum communication, the actual available W states tend to be non-maximal ones due to environmental noise and operational imperfections. Furthermore, it has been demonstrated that compared with the standard W state, the so-called W-like state can be used to achieve perfect teleportation and perfect superdense coding \cite{gorbachev2003can,PhysRevA.74.062320,li2007states}. Therefore, it is of great significance to prepare or fuse such W-like states.

In this paper, we use the cavity QED system to realize the fusion of W-like states. The cavity QED system is one of the well realization platforms for QIP. Using the large-detuning interaction between atoms and a single-mode cavity field \cite{PhysRevA.65.042102}, we realize two fusion schemes for W-like states, i.e., to fusing $|\mathcal{W}_N\rangle$ and $|\mathcal{W}_M\rangle$ into a large-scare $|\mathcal{W}_{N+M-1}\rangle$, and a $|\mathcal{W}_{N+M+T-2}\rangle$ from $|\mathcal{W}_N\rangle$, $|\mathcal{W}_M\rangle$ and $|\mathcal{W}_T\rangle$. Moreover, we calculate the success probabilities for these two schemes, respectively, which are larger than that of a similar optical fusion scheme \cite{li2016generating}. Thus, we numerically compare their resource costs, and the results show that our scheme is better. In addition, due to the influence of environmental noise, we take the decay and pure dephasing of the atoms into account, and the numerical simulation shows that our scheme is feasible. Note that the single-mode cavity involved is always in the vacuum state during the fusion process, so we do not need to consider the decay of the cavity mode. Besides, our fusion mechanism is based on the large-detuning interaction between a single-mode cavity and several atoms, which has been experimentally demonstrated \cite{PhysRevLett.87.037902}. In summary, our fusion scheme is very likely to be realized experimentally. The paper is arranged as follows. In the Sec. \ref{sec2}, we give the fusion mechanism and fusion scheme for two small-scale atomic W-likes; As a generalization, in the Sec. \ref{sec3}, we give a scheme of fusing three atomic W-like states into a large-scale atomic W-like state; The next section is the analysis and discussion, including the comparison of resource costs of different schemes and the experimental feasibility analysis considering decoherence. Finally, the conclusion of the paper is given.

\section{Fusion scheme of two small-size entangled atomic W-like states}\label{sec2}
\subsection{Fusion mechanism}
Before proceeding to the specific fusion schemes, let's first introduce the fusion mechanism, which is based on the large detuning interaction of atoms with a single-mode cavity field. Specifically, suppose that there are $k$ identical two-level atoms interacting with a single-mode cavity field. In the interaction picture, its interaction Hamiltonian \cite{PhysRevA.65.042102} is
\begin{equation}
H_I=\mathrm{g}\sum_{j=1}^{k}\left(e^{-i\Delta t}a^{\dagger}s^{-}_{j}+e^{i\Delta t}a s_{j}^{+}    \right),
\end{equation}
where $s^{+}_{j}=|e_j\rangle\langle g_j|$, $s^{-}_{j}=|g_j\rangle\langle e_j|$, $|e_j\rangle$ and $|g_j\rangle$ denote the excited and ground states of the $j$-th atom, respectively. Besides, $a^{\dagger}$ and $a$ denote the generation and annihilation operators of the cavity field, respectively, $\mathrm{g}$ is the coupling constant between each atom and the cavity mode, and $\Delta$ denotes the detuning between the atomic transition frequency and the frequency of cavity field. When $\Delta\gg\mathrm{g}$, there is no energy exchange between the cavity mode and the atoms, whose effective Hamiltonian can be written as
\begin{equation}
H_{e}=\lambda\left[\sum_{i,j=1}^{k}(s^{+}_{j}s^{-}_{i}a a^{\dagger}-s^{-}_{j}s^{+}_{i}a^{\dagger}a)  \right],
\end{equation}
in which $\lambda=\mathrm{g}^2/\Delta$. Particularly, when the cavity mode is initially prepared in the vacuum state $|0\rangle$, the above equation will reduce to
\begin{equation}\label{eq3}
H_{e}'=\lambda\left(\sum_{j=1}^{k}|e\rangle_{jj}\langle e|+\sum_{i,j=1,i\neq j}^{k}s^{+}_{j}s^{-}_{i}  \right),
\end{equation}
which only involves the interaction between atoms, and the cavity mode is virtually excited. We can let $k=2$, then Eq. (\ref{eq3}) becomes $H_{e}'=\lambda\left[\sum_{j=1}^{2}|e\rangle_{jj}\langle e|+(s_1^+ s_2^-+s_1^- s_2^+)  \right]$. Substituting it into the Schrodinger equation $i\hbar |\dot{\psi}\rangle=H_e'|\psi\rangle$, one can solve them for several different initial states, as
\begin{eqnarray}\label{eq4}
\begin{split}
|ee\rangle_{12}&\longrightarrow e^{-i2\lambda t}|ee\rangle_{12}, \\
|eg\rangle_{12}&\longrightarrow e^{-i\lambda t}\left(\cos{\lambda t|eg\rangle_{12}-i\sin{\lambda t}|ge\rangle_{12}} \right), \\
|ge\rangle_{12}&\longrightarrow e^{-i\lambda t}\left(\cos{\lambda t|ge\rangle_{12}-i\sin{\lambda t}|eg\rangle_{12}} \right), \\
|gg\rangle_{12}&\longrightarrow |gg\rangle_{12}.
\end{split}
\end{eqnarray}
Similarly, for $k=3$, we can also obtain the following evolutions for the different initial states:
\begin{eqnarray}\label{eq5}
|eee\rangle_{123}&\longrightarrow& e^{-i3\lambda t}|eee\rangle_{123},  \nonumber \\
|gee\rangle_{123}&\longrightarrow& e^{-i\lambda t}(B|gee\rangle_{123}+A|ege\rangle_{123}+A|eeg\rangle_{123}),  \nonumber\\
|ege\rangle_{123}&\longrightarrow& e^{-i\lambda t}(A|gee\rangle_{123}+B|ege\rangle_{123}+A|eeg\rangle_{123}),  \nonumber\\
|eeg\rangle_{123}&\longrightarrow& e^{-i\lambda t}(A|gee\rangle_{123}+A|ege\rangle_{123}+B|eeg\rangle_{123}), \nonumber\\
|gge\rangle_{123}&\longrightarrow& (A|egg\rangle_{123}+A|geg\rangle_{123}+B|gge\rangle_{123}), \nonumber\\
|geg\rangle_{123}&\longrightarrow& (A|egg\rangle_{123}+B|geg\rangle_{123}+A|gge\rangle_{123}), \nonumber\\
|egg\rangle_{123}&\longrightarrow& (B|egg\rangle_{123}+A|geg\rangle_{123}+A|gge\rangle_{123}), \nonumber\\
|ggg\rangle_{123}&\longrightarrow& |ggg\rangle_{123},
\end{eqnarray}
where $A=(e^{-i3\lambda t}-1)/3$, $B=(e^{-i3\lambda t}+2)/3$.

\begin{figure}[tbp]
\centering
\includegraphics[width=0.75\linewidth]{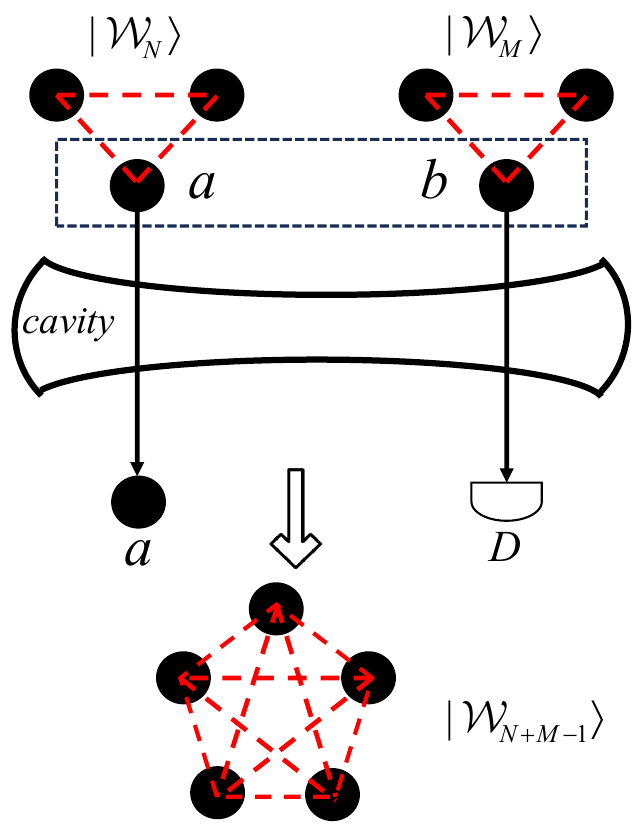}
\caption{The setup for fusing two small-scale atomic W-like states into a large-scale one in cavity QED system, where $D$ denotes an atomic detector.}\label{Figure1}
\end{figure}

\subsection{Fusion scheme for two small-scale atomic W-likes}
As shown in Fig. \ref{Figure1}, we extract one atom $a$ and one atom $b$ from an $N$-atom W-like state $|\mathcal{W}_N\rangle$ and an $M$-atom W-like state $|\mathcal{W}_M\rangle$, respectively, and simultaneously send these two atoms into the large-detuning cavity with single mode. When these two atoms fly out of the cavity, we can use atomic detector $D$ to detect the state of any one of them (selecting atom $b$ here). Based on different detection results, we can determine whether the fusion process is successful or not. If the state of atom $b$ is detected to be in excited state $|e\rangle_b$, the fusion is successful. That is to say, an $N$-atom W-like state $|\mathcal{W}_N\rangle$ and an $M$-atom W-like state $|\mathcal{W}_M\rangle$ are fused into an $(N+M-1)$-atom W-like state $|\mathcal{W}_{N+M-1}\rangle$, otherwise, the fusion is fail. The specific process is as follows:

Firstly, $N$-atom W-like state and $M$-atom W-like state can be expressed as
\begin{eqnarray}
\begin{split}
|\mathcal{W}_N\rangle&=\frac{1}{\sqrt{2}}\left[|gg...g\rangle|e\rangle_a+|W_{N-1}\rangle|g\rangle_a\right], \\
|\mathcal{W}_M\rangle&=\frac{1}{\sqrt{2}}\left[|gg...g\rangle|e\rangle_b+|W_{M-1}\rangle|g\rangle_b\right],
\end{split}
\end{eqnarray}
where $|g\rangle$ and $|e\rangle$ indicate that the atom is in the ground and excited states, respectively, and $|W_{N-1}\rangle$ (or $|W_{M-1}\rangle$) is the standard entangled W state, which can be read as $|W_{N-1}\rangle=\left[|gg...g\rangle|e\rangle+\sqrt{N-2}|W_{N-2}\rangle|g\rangle\right]/\sqrt{N-1}$.

Suppose that there are two small-scale atomic W-like states $|\mathcal{W}_N\rangle$ and $|\mathcal{W}_M\rangle$, our goal is to merge these two small-scale atomic W-like states into a larger W-like state $|\mathcal{W}_{M+N-1}\rangle$. Thus, we will simultaneously send two atoms $a$ and $b$ from these two small-scale W-like states into a detuned cavity, and a single cavity mode will induce interaction between these two atoms. The initial state of the entire system can be represented as:
\begin{eqnarray}
\begin{split}
|\psi_1(0)\rangle&=|\mathcal{W}_N\rangle\otimes|\mathcal{W}_M\rangle=\frac{1}{2}[|gg...g\rangle|gg...g\rangle|ee\rangle_{ab}    \\
&+|gg...g\rangle\otimes|W_{M-1}\rangle|eg\rangle_{ab}+|W_{N-1}\rangle|ggg...\rangle  \\
&\otimes|ge\rangle_{ab}+|W_{N-1}\rangle|W_{M-1}\rangle|gg\rangle_{ab}].
\end{split}
\end{eqnarray}
According to Eq. (\ref{eq4}), the state of the entire system will become as for time $t$ after these atoms fly out of the cavity:
\begin{widetext}
\begin{eqnarray}
\begin{split}
|\psi_1(t)\rangle&=\frac{1}{2}\Big[e^{-i2\lambda t}|gg...g\rangle|gg...g\rangle|ee\rangle_{ab}+e^{-i\lambda t}(\cos{\lambda t}|gg...g\rangle|W_{M-1}\rangle-i\sin{\lambda t}|W_{N-1}\rangle|gg...g\rangle)|eg\rangle_{ab}\\
&\quad+e^{-i\lambda t}(\cos{\lambda t}|W_{N-1}\rangle|gg...g\rangle-i\sin{\lambda t} |gg...g\rangle|W_{M-1}\rangle)|ge\rangle_{ab}+|W_{N-1}\rangle|W_{M-1}\rangle|gg\rangle_{ab}\Big].
\end{split}
\end{eqnarray}
\end{widetext}
When the result of atomic detector $D$ is $|e\rangle_b$, the remaining part of the system will collapse to the following state:

\begin{widetext}
\begin{equation}\label{eq44}
\begin{split}
|\psi_1(t)\rangle_{N+M-1}=\frac{1}{2}\Big[e^{-i2\lambda t}|gg...g\rangle|gg...g\rangle|e\rangle_{a}+e^{-i\lambda t}(\cos{\lambda t}
|W_{N-1}\rangle|gg...g\rangle-i\sin{\lambda t} |gg...g\rangle|W_{M-1}\rangle)|g\rangle_{a}\Big].
\end{split}
\end{equation}
\end{widetext}

If we let $|\frac{\cos{\lambda t}}{\sqrt{N-1}}|=|\frac{\sin{\lambda t}}{\sqrt{M-1}}|=\frac{1}{\sqrt{M+N-2}}$, then the Eq. (\ref{eq44}) becomes
\begin{widetext}
\begin{equation} \label{eq55}
\begin{split}
|\psi_1 \rangle_{N+M-1}=\frac{1}{2}\Big[e^{-i\lambda t}|gg...g\rangle|gg...g\rangle|e\rangle_{a}+ \frac{\sqrt{N-1}}{\sqrt{N+M-2}}|W_{N-1}\rangle|gg...g\rangle|g\rangle_a-i\frac{\sqrt{M-1}}{\sqrt{N+M-2}} |gg...g\rangle|W_{M-1}\rangle|g\rangle_{a}\Big],
\end{split}
\end{equation}
\end{widetext}
in which the global phase factor $e^{-i\lambda t}$ has been omitted. If we make several the same single-qubit local unitary operations on the $N$-th to $(N+M-2)$-th atoms, respectively, and a phase gate on the $(N+M-1)$-th atom for Eq. (\ref{eq55}), we can convert them into the standard $(N+M-1)$-atom W-like state, as
\begin{equation}
|\psi'_1\rangle_{N+M-1}=\frac{1}{\sqrt{2}}|\mathcal{W}_{N+M-1}\rangle.
\end{equation}
According to the above equation, the success probability of fusion process is $1/2$. However, for different $N$ and $M$ values, the interaction time between the fused atoms and the single-mode cavity field is changing, which can be divided into two categories. Note that it is assumed $\lambda t\in\left(0,\frac{\pi}{2}\right)$ here, which is specifically as follows:

(1) when $N=M$, the solution yields $\lambda t=\frac{\pi}{4}$;

(2) when $N\neq M$, we have
\begin{equation}
\left|\frac{\cos{\lambda t}}{\sqrt{N-1}}\right|=\left|\frac{\sin{\lambda t}}{\sqrt{M-1}}\right|,
\end{equation}
and then it can obtain that a series of different $\lambda t$ values based on the different choice of $N$ and $M$. To sum up, we implement the fusion of two small-size atomic W-like states into one large-size atomic W-like state in the cavity QED systems.

\begin{figure}[tbp]
\centering
\includegraphics[width=0.80\linewidth]{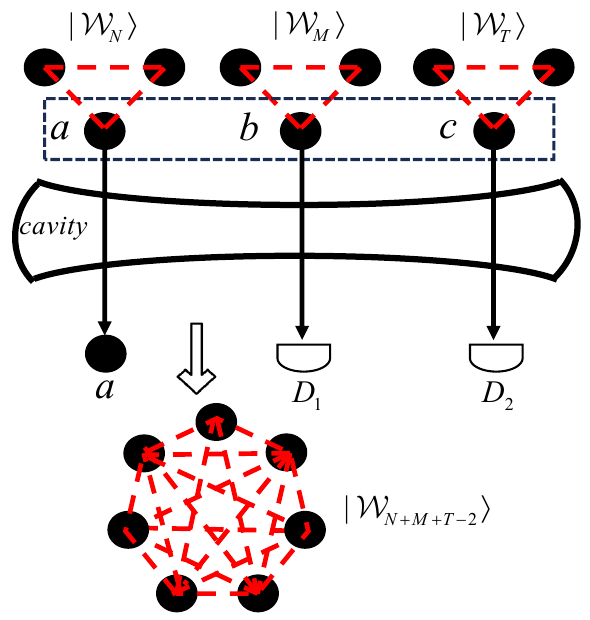}
\caption{The setup for fusing three small-scale atomic W-like states into a large-scale one in cavity QED system, where $D_1$ and $D_2$ are two atomic detectors.}\label{Figure2}
\end{figure}
\section{Fusion scheme of three small-size entangled atomic W-like states}\label{sec3}
Similarly, as shown in Fig. \ref{Figure2}, firstly, we extract one atom to be fused from each of the three small-scale atomic W-like states $|\mathcal{W}_N\rangle$, $|\mathcal{W}_M\rangle$ and $|\mathcal{W}_T\rangle$, respectively, denoted as $a$, $b$, $c$, and simultaneously send these three atoms into a single-mode cavity to undergo a large detuning interaction with the cavity field. The initial state of the entire system can be represented in the following form:
\begin{widetext}
\begin{eqnarray}
\begin{split}
|\psi_2(0)\rangle&=|\mathcal{W}_N\rangle\otimes|\mathcal{W}_M\rangle\otimes|\mathcal{W}_T\rangle=\frac{1}{2\sqrt{2}}\Big[|gg...g\rangle|gg...g\rangle|gg...g\rangle|eee\rangle_{abc}
+|W_{N-1}\rangle|gg...g\rangle|gg...g\rangle|gee\rangle_{abc}\\
&\quad+|gg...g\rangle|W_{M-1}\rangle|gg...g\rangle|ege\rangle_{abc}+|gg...g\rangle|gg...g\rangle|W_{T-1}\rangle|eeg\rangle_{abc}+|W_{N-1}\rangle|W_{M-1}\rangle|gg...g\rangle|gge\rangle_{abc}\\
&\quad+|W_{N-1}\rangle|gg...g\rangle|W_{T-1}\rangle|geg\rangle_{abc}+|gg...g\rangle|W_{M-1}\rangle|W_{T-1}\rangle|egg\rangle_{abc}+|W_{N-1}\rangle|W_{M-1}\rangle|W_{T-1}\rangle|ggg\rangle_{abc}\Big].
\end{split}
\end{eqnarray}
\end{widetext}
According to the Eq. (\ref{eq5}) for the large-detuning interaction between three atoms and single-mode cavity, the quantum state of the entire system at time $t$ will evolve to
\begin{widetext}
\begin{eqnarray}
\begin{split}
|\psi_2(t)\rangle&=\frac{1}{2\sqrt{2}}\Big[|gg...g\rangle|gg...g\rangle|gg...g\rangle e^{-i3\lambda t}|eee\rangle_{abc}
+|W_{N-1}\rangle|gg...g\rangle|gg...g\rangle e^{-i\lambda t}\big(B|gee\rangle_{abc}+A|ege\rangle_{abc}\\
&\quad+A|eeg\rangle_{abc}\big) +|gg...g\rangle|W_{M-1}\rangle|gg...g\rangle e^{-i\lambda t}\big(A|gee\rangle_{abc}+B|ege\rangle_{abc}+A|eeg\rangle_{abc}\big)+|gg...g\rangle|gg...g\rangle \\
&\quad\otimes|W_{T-1}\rangle e^{-i\lambda t}\big(A|gee\rangle_{abc}+A|ege\rangle_{abc}+B|eeg\rangle_{abc}\big)+|W_{N-1}\rangle|W_{M-1}\rangle|gg...g\rangle(A|egg\rangle_{abc}+A|geg\rangle_{abc}\\
&\quad+B|gge\rangle_{abc})+|W_{N-1}\rangle|gg...g\rangle|W_{T-1}\rangle\big(A|egg\rangle_{abc}+B|geg\rangle_{abc}+A|gge\rangle_{abc}\big)+|gg...g\rangle|W_{M-1}\rangle|W_{T-1}\rangle\\
&\quad\otimes\big(B|egg\rangle_{abc}+A|geg\rangle_{abc}+A|gge\rangle_{abc}\big)+|W_{N-1}\rangle|W_{M-1}\rangle|W_{T-1}\rangle|ggg\rangle_{abc}\Big].
\end{split}
\end{eqnarray}
\end{widetext}

By utilizing atomic detectors $D_1$ and $D_2$, respectively, to measure any two of the three atoms that fly out of the single-mode cavity (selecting atoms $b$ and $c$ here), we can determine whether the fusion process is successful or not. It can be shown that is positive when the composite state of atoms $b$ and $c$ is measured to be $|ee\rangle_{bc}$. At this point, the quantum state of the entire system will collapse to the following form:
\begin{widetext}
\begin{eqnarray}\label{eq10}
\begin{split}
|\psi_2\rangle_{N+M+T-2}&=\frac{1}{2\sqrt{2}}\Big[|gg...g\rangle|gg...g\rangle|gg...g\rangle e^{-i3\lambda t}|e\rangle_a+|W_{N-1}\rangle|gg...g\rangle|gg...g\rangle e^{-i\lambda t}B|g\rangle_a \\
&\quad+|gg...g\rangle|W_{M-1}\rangle|gg...g\rangle e^{-i\lambda t}A|g\rangle_a+|gg...g\rangle|gg...g\rangle|W_{T-1}\rangle e^{-i\lambda t}A|g\rangle_a\Big].
\end{split}
\end{eqnarray}
\end{widetext}

When it can be satisfied
\begin{equation}
\left|\frac{B}{\sqrt{N-1}}\right|=\left|\frac{A}{\sqrt{M-1}}\right|=\left|\frac{A}{\sqrt{T-1}}\right|=\frac{1}{\sqrt{N+M+T-3}}, \nonumber
\end{equation}
then Eq. (\ref{eq10}) will become a non-standard  $(N+M+T-2)$-atomic W-like state, which can use classical optical pulses to eliminate relative phase and obtain the standard W-like states:
\begin{equation}
|\psi'_2\rangle_{N+M+T-2}=\frac{1}{2}|\mathcal{W}_{N+M+T-2}\rangle.
\end{equation}
As can be seen from the above equation, it will be eventually fused into an $(N+M+T-2)$-atomic W-like state from three small-size atomic W-like state with a $1/4$ probability. Similarly, the interaction time can also be divided into two situations, as discussed below:

(1) if $N=M=T$, the solution is $\lambda t=\frac{2\pi}{9}$;

(2) if $N\neq M=T$, one must be satisfied
\begin{equation}
\left|\frac{B}{\sqrt{N-1}}\right|=\left|\frac{A}{\sqrt{M-1}}\right|,
\end{equation}
which indicates that we can obtain a series of different $\lambda t$ values for different choice of $N$, $M$ and $T$. It is worth noting that $M$ and $T$ must be equal here.
In summary, we achieve the fusion of three small-size atomic W-like states into a large-size atomic W-like state in the cavity QED systems.

In principle, our fusion scheme can be extended to the fusion of $k$ small-scale W-like states based on the Eq. (\ref{eq3}) with $k\geq4$. Moreover, for each individual fusion scheme, the success probability of that is still fixed regardless of the scale of the fused W-like states, i.e., $P_s(\mathcal{W}_N^1,\mathcal{W}_M^2,...,\mathcal{W}_Z^k)=1/2^{k-1}$. However, for most of the current fusion schemes of W or W-like states, their success probability of fusion decreases as the scale of the target large-scale W or W-like states increases, which may results in low fusion efficiency.

\section{Analysis and discussion}
\begin{figure}[tbp]
\centering
\includegraphics[width=0.90\linewidth]{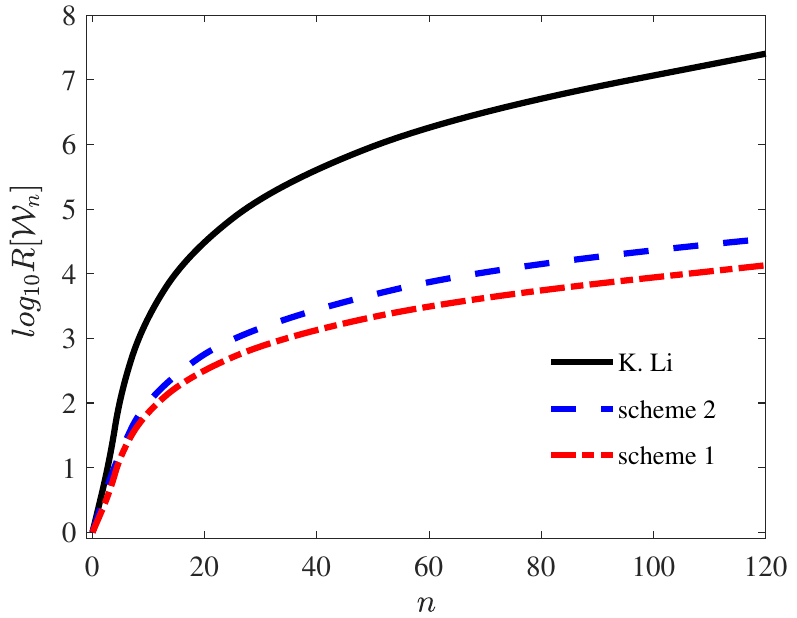}
\caption{Comparison of optimal resource cost.  The horizontal axis indicates the scale $n$ of the target W-like state $|\mathcal{W}_n\rangle$, and the vertical axis indicates the corresponding resource cost $\log_{10}R[\mathcal{W}_n]$. The black-solid line indicates the fusion of two small-scale photonic W-like states in Ref. \cite{li2016generating}, while the red dot-dashed and blue-dashed lines represent the fusion of two and three small-scale atomic W-like states in this paper, respectively.}\label{Figure3}
\end{figure}

From the fusion schemes for two kind of atomic W-like states in the Sec. \ref{sec2} and Sec. \ref{sec3}, it can be found that the success probability of fusion is fixed for different $N$, $M$, or $N$, $M$ and $T$ values, which is different from the existing fusion schemes for atomic W and photonic W or W-like states \cite{ozdemir2011optical,PhysRevA.87.032331,li2016generating,zang2017generating,li2018preparing,PhysRevA.94.062315,zang2015generating,ding2018qubit,shao2024utilizing,han2017effective,huang2024quantum} In these fusion schemes, the success probability of fusion will sharply decrease with the increase of particle numbers of target states, which will inevitably lead to a considerable increase in resource cost. Thus, to highlight the advantages of our scheme in resource cost, we evaluate this scheme using the optimal resource cost method \cite{ozdemir2011optical} and compare it with the fusion scheme for photonic W-like states in Ref. \cite{li2016generating}, as shown in Fig. 3.  The resource cost for generating a $|\mathcal{W}_{N+M-1}\rangle$ from a $|\mathcal{W}_{N}\rangle$ and a $|\mathcal{W}_{M}\rangle$ can be defined as
\begin{equation}
R[\mathcal{W}_{N+M-1}]=\frac{R[\mathcal{W}_{N}]+R[\mathcal{W}_{M}]}{P_s(\mathcal{W}_{N},\mathcal{W}_{M})},
\end{equation}
in which $P_s(\mathcal{W}_{N},\mathcal{W}_{M})$ denotes the success probability for our fusion process, which is equal to $1/2$ for different $N$ and $M$. Herein, we set $R[\mathcal{W}_2]=1$ as the basic unit of resource cost, and choose the strategy of optimal resource cost for comparison. In addition, the definition of resource cost for fusing three small-scale W-like states is the similar as that of $|\mathcal{W}_{N+M-1}\rangle$. From the Fig. \ref{Figure3}, it can be observed that the resource costs of both our fusion schemes for two kinds of atomic W-like states are lower than that of scheme for the photonic W-like state in Ref. \cite{li2016generating}, which shows our schemes are more efficient and feasible.

In addition, although the success probabilities of our fusion schemes are fixed, the interaction time between atoms and cavity will vary with different $N$, $M$, and $T$ values. That is, to ensure successful fusion, we need to precisely control the flight speed of atoms. Meanwhile, due to the dissipative effect between the cavity and atoms, it is necessary to consider whether this scheme still has high feasibility and practicality in the presence of decoherence.

\setlength{\tabcolsep}{10mm}{\begin{table*}
\centering
\caption{The required times $\lambda t_1$ and $\lambda t_2$ for fusing two and three small-scale atomic W-like states,
respectively, for different input scales of $N$, $M$, and $N$, $M$, $T$.}\label{Table}
\begin{tabular}{ccc|cccc}
\hline\hline\noalign{\smallskip}
$N$ & $M$&$\lambda t_1$&$N$&$M$&$T$&$\lambda t_2$ \\
\noalign{\smallskip}\hline\noalign{\smallskip}
2 & 2 &  0.7854 &2 &2&2&0.6981\\
2 & 3 &  0.9553 &4&2&2&0.4902 \\
3 & 2  & 0.6155 &2&4&4&0.9204 \\
3 & 3  & 0.7854 &4&4&4&0.6981 \\
3 & 4  & 0.8861 &6&4&4&0.6000 \\
4 & 3  & 0.6847 &4&6&6&0.7967 \\
4 & 4  & 0.7854 &6&6&6&0.6981 \\
4 & 5  & 0.8571 &8&6&6&0.6334 \\
5 & 4  & 0.7137 &6&8&8&0.7629 \\
5 & 5  & 0.7854 &10&10&10&0.6981 \\
\noalign{\smallskip}\toprule
\end{tabular}
\end{table*}}

\begin{figure}[tbp]
\centering
\includegraphics[width=0.90\linewidth]{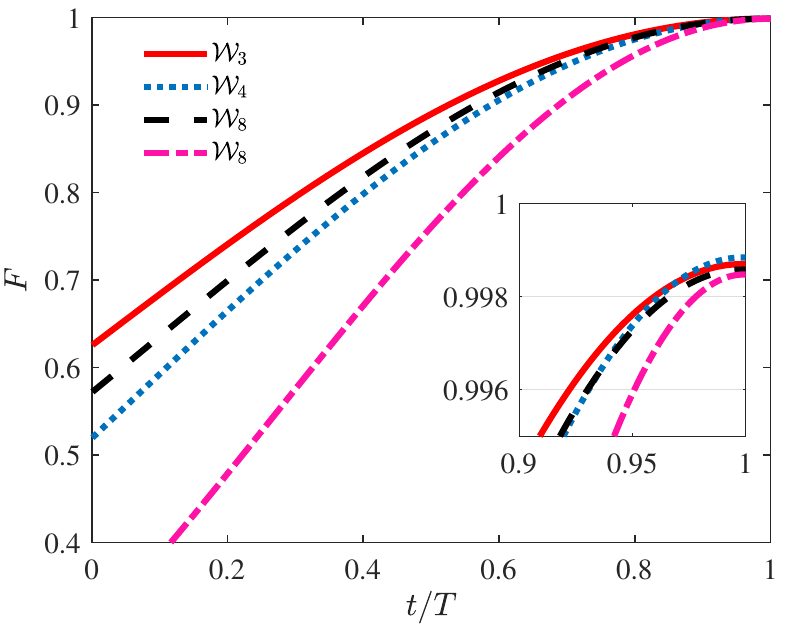}
\caption{Fidelity $F$ of the target atomic W-like state $|\mathcal{W}_n\rangle$ using different fusion schemes in the paper as a function of evolution time $t$, in which $n$ takes $3$, $4$ and $8$, respectively. The red-solid line represents the target $|\mathcal{W}_3\rangle$ from two $|\mathcal{W}_2\rangle$; The blue-dotted line indicates the target  $|\mathcal{W}_4\rangle$ from three $|\mathcal{W}_2\rangle$; The target $|\mathcal{W}_8\rangle$ can be fused from a $|\mathcal{W}_4\rangle$ and a $|\mathcal{W}_5\rangle$ (black-dashed line), or a $|\mathcal{W}_2\rangle$ and two $|\mathcal{W}_4\rangle$ (pink dot-dashed line). }\label{Figure4}
\end{figure}

Firstly, we give the required time of the fusion process of fusing two or three atomic W-like states  for different $N$, $M$, or $N$, $M$, $T$ values, as shown in Table \ref{Table}. From the Table \ref{Table}, it can be seen that no matter how large the values of $N$, $M$, or $N$, $M$, $T$ are, there is always $\lambda t<\pi/2$, and the interaction times between the atoms and the cavity mode are $t<\pi/(2\lambda)$. If we use Rydberg atoms with principal quantum numbers $49$, $50$, and $51$, then the coupling constant is $\mathrm{g}=2\pi\times24$ kHz \cite{PhysRevLett.77.4887}. Here we take the detuning being $\Delta=10\mathrm{g}$, so $t$ is about on the order of $10^{-4}$s, Furthermore, the decay time of typical atoms is approximately $T_r=3\times10^{-2}$s \cite{PhysRevLett.87.037902}, and we utilize a cavity system with a quality factor of $Q=10^8$ \cite{PhysRevLett.77.4887}, whose decay time of the cavity mode is about $T_c=3\times10^{-2}$s. Of course, the used cavity mode is initially prepared on a vacuum state and is only virtually excited throughout the evolution, so cavity leakage is greatly suppressed. In Fig. \ref{Figure4}, we numerically simulate the dynamics of the target W-like states using the Lindblad quantum master equation \cite{lindblad1976generators} of
\begin{equation}
\dot{\rho}=-i[H_{e}',\rho]+\frac{\kappa_{-}}{2}\mathcal{L}(\sigma_{-})+\frac{\kappa_{z}}{2}\mathcal{L}(\sigma_{z}),
\end{equation}
where $\rho$ indicates the density operator of the total quantum system,  the Lindblad operator $\mathcal{L}(\sigma)$ is $2\sigma\rho\sigma^{\dagger}-\sigma^{\dagger}\sigma\rho-\rho\sigma^{\dagger}\sigma$ for operator $\sigma$, $\sigma_{-}=|g\rangle\langle e|$, $\sigma_{z}=|e\rangle\langle e|-|g\rangle\langle g|$, and $\kappa_{-}$ and $\kappa_{z}$ are atomic decay and dephasing rates, respectively. We define the fidelity $F=\mathrm{Tr}\left(\rho|\mathcal{W}_n\rangle\langle\mathcal{W}_n|\right)$ with the ideal target W-like states $|\mathcal{W}_n\rangle$. Here, we take $n=3, 4, 8$ as an example, that is, using the first scheme, two $|\mathcal{W}_2\rangle$ states can be fused into a $|\mathcal{W}_3\rangle$ state and a $|\mathcal{W}_8\rangle$ state is fused from a $|\mathcal{W}_4\rangle$ and a $|\mathcal{W}_5\rangle$ states. On the contrary, the target $|\mathcal{W}_4\rangle$ and a same $|\mathcal{W}_8\rangle$ states can be fused from three $|\mathcal{W}_2\rangle$ states, and a $|\mathcal{W}_2\rangle$ and two $|\mathcal{W}_5\rangle$ states using the second scheme, respectively. As shown in Fig. \ref{Figure4}, the fidelities of target W-like states $|\mathcal{W}_3\rangle$, $|\mathcal{W}_4\rangle$, $|\mathcal{W}_8\rangle$ using our different  fusion strategies can all reach above $99.8\%$, in which we only consider decay and dephasing rates of atoms $\kappa_{-}=\kappa_z=1/T_r$.
To sum up, it is not difficult to find that our fusion schemes are possible to be implemented under the current experimental conditions.

\section{Conclusion}
In summary, we propose two kinds of fusion schemes of atomic W-like states via detecting the atomic states, which can achieve the quantum fusion of  two atomic W-like states $|\mathcal{W}_{N}\rangle$ and $|\mathcal{W}_{M}\rangle$ to a large-scale atomic W-like state $|\mathcal{W}_{N+M-1}\rangle$, and three atomic W-like states $|\mathcal{W}_{N}\rangle$, $|\mathcal{W}_{M}\rangle$, $|\mathcal{W}_{T}\rangle$ to a large-scale atomic W-like state $|\mathcal{W}_{N+M+T-2}\rangle$. The success probability of our fusion schemes is high and fixed, regardless of the scale of the target atomic W-like states, which can induce high fusion efficiency. In addition, the resource cost and feasibility analysis indicate that this scheme is feasible under the current experimental conditions. Our study may serve as a foundation for the development of perfect teleportation and densecoding based on large-scale atomic W-like states.

\begin{acknowledgements}
This work is supported by the National Natural Science Foundation of China(No.12204013), the key Scientific Research Foundation of Anhui Provincial Education Department (No.2023AH050481), the Natural Science Foundation of Anhui Province(No.1708085MA10), the Quality Engineering
Project of Anhui Provincial Education Department (No.2023zygzts036), and the Science Foundation in Colleges and Universities of Anhui Province of China (No. KJ2019A0562).

\end{acknowledgements}

%\bibliographystyle{apsrev4-1}
%\bibliography{ref}%¼ÓÈë²Î¿¼ÎÄÏ×
%

\end{document}